\documentclass[aps,twocolumn,letterpaper,longbibliography]{revtex4-1}
\usepackage{amssymb}
\usepackage{amsmath}
\usepackage{graphicx}
\usepackage{extarrows}
\usepackage{url}
\usepackage{varioref}
\usepackage{hyperref}
\usepackage{cleveref}

\usepackage{xcolor}
\usepackage[normalem]{ulem}

\usepackage{verbatim} %Liangchao

\begin{document}

\title{Dynamics of domain walls in a Bose-Einstein condensate driven by density-dependent gauge field} 

% Place the author information here.  Please hand-code the contact
% information and notecalls; do *not* use \footnote commands.  Let the
% author contact information appear immediately below the author names
% as shown.  We would also prefer that you don't change the type-size
% settings shown here.
\author{Kai-Xuan Yao}
\author{Zhendong Zhang}
\author{Cheng Chin}

\affiliation{James Franck Institute, Enrico Fermi Institute and Department of Physics, University of Chicago, Chicago, Illinois 60637, USA}

\begin{abstract}
Dynamical coupling between matter and gauge fields underlies the emergence of many exotic particle-like excitations in condensed matter and high energy physics.
% Quantum simulation of dynamical gauge fields promises a powerful tool to study many-body phenomena relevant to many areas in condensed matter and high energy physics.
% One example is the formation of exotic particle-like excitations as a result of the matter-gauge interaction.
An important stepping stone to simulate this physics in atomic quantum gases relies on the synthesis of density-dependent gauge fields.
Here we demonstrate deterministic formation of domain walls in a stable Bose-Einstein condensate with a synthetic gauge field that depends on the atomic density.
% Here we experimentally prepare a Bose-Einstein condensate (BEC) in the presence of a density-dependent gauge field, and observe the resulting formation and dynamics of domain walls separating regions with different momenta.
The gauge field is created by simultaneous modulations of the optical lattice potential and interatomic interactions, and results in domains of atoms condensed into two different momenta.
% We engineer a gauge field that takes one of two values depending on the density, by combining two-frequency shaking of a 1D optical lattice and modulation of the scattering length.
% Preparing a BEC in the presence of the density-dependent gauge field, we observe formation of domains whose momentum depends on the local density.
% In response to a synthetic electric field, we observe the res
Modeling the domain walls as elementary excitations, we find that the domain walls respond to synthetic electric field with a charge-to-mass ratio larger than and opposite to that of the bare atoms.
% We investigate the dynamics of the domain walls in response to a synthetic electric field, and observe a response larger than and opposite to that of the atoms.
% , and find that the response is drastically different from the bare atoms, in the opposite direction and with larger magnitude.
Our work offers promising prospects to simulate the dynamics and interactions of novel excitations in quantum systems with dynamical gauge fields.
% Our work constitutes the first experimental investigation of many-body physics in the presence of density-dependent gauge field, and presents a platform for studying the dynamical properties of topological defects in a quantum many-body system.
\end{abstract}

% Include the date command, but leave its argument blank.

%\date{}

%%%%%%%%%%%%%%%%% END OF PREAMBLE %%%%%%%%%%%%%%%%

%\usepackage{multicol,lipsum}
%\begin{document} 

% Double-space the manuscript.

%\baselineskip24pt

% Make the title.

\maketitle 

Gauge theories form a cornerstone in our understanding of condensed matter systems \cite{Kogut1979} and fundamental particles \cite{Wilson1974}.
A complete theoretical understanding of many-body systems subject to gauge fields, however, faces significant analytical and numerical challenges \cite{Alford2008,Troyer2005}.
Experiments with ultracold atoms offer an alternative approach by quantum simulating gauge theory models, where gauge fields can be artificially synthesized \cite{Goldman2014,Zohar2016,Cooper2019}. Tremendous progress has been made in the past years on creating static artificial gauge fields in atomic quantum gases \cite{Spielman2009}, enabling the realization of, for instance, the iconic Haldane \cite{Esslinger2014} and Hofstadter models \cite{Bloch2013, Ketterle2013}.
Fundamentally, gauge fields are dynamical with quantum degrees of freedom that interact with matter \cite{Baskaran1988,Cheng1994,Levin2005,Wiese2013,Savary2016}.
An intriguing consequence of the dynamical feedback between the matter and gauge field is the formation of novel particle-like excitations with emergent properties, for example, mesons in the standard model \cite{Griffiths2008} and composite fermions in the fractional quantum Hall effect \cite{Stormer1999}.
% Many important phenomena in condensed matter and high energy physics are described by gauge fields that are intrinsically dynamical \cite{Cheng1994,Levin2005,Savary2016}. 
% These gauge fields are quantum degrees of freedom that interact with matter, and the back-action of matter onto the gauge fields is a crucial ingredient.
% Recently, several experiment groups have made progress in simulating dynamical gauge fields, by realizing lattice gauge theory models \cite{Schweizer2019,Yang2020}
% An important stepping stone to access this exciting physics is the creation of density-dependent gauge fields, whose configuration depends on the density distribution of matter \cite{Ohberg2013}.
% These systems are predicted to host a number of new many-body phenomena, including flux attachment \cite{Ohberg2020}, phase transitions induced by exchange statistics \cite{Keilmann2011}, geometric frustration \cite{Mishra2016} and quantum scarring \cite{Hudomal2020}.
Recently, several experiment groups have realized density-dependent gauge fields \cite{Clark2018,Gorg2019,Lienhard2020}, where the strength of the field depends on the density of matter \cite{Ohberg2013}, as well as lattice gauge theory models \cite{Schweizer2019,Yang2020,Mil2020}.

% 1. broader physics from dynamical gauge field, theory predictions (general dynamical gauge theories)

% 2. explain what dynamical gauge field means, crucial ingredients

% 3. one stepping stone (example) is DDG, cite some theory papers on DDG

% 4. experiment achievements (4 papers)

\begin{figure}[t]
\includegraphics[width=86mm]{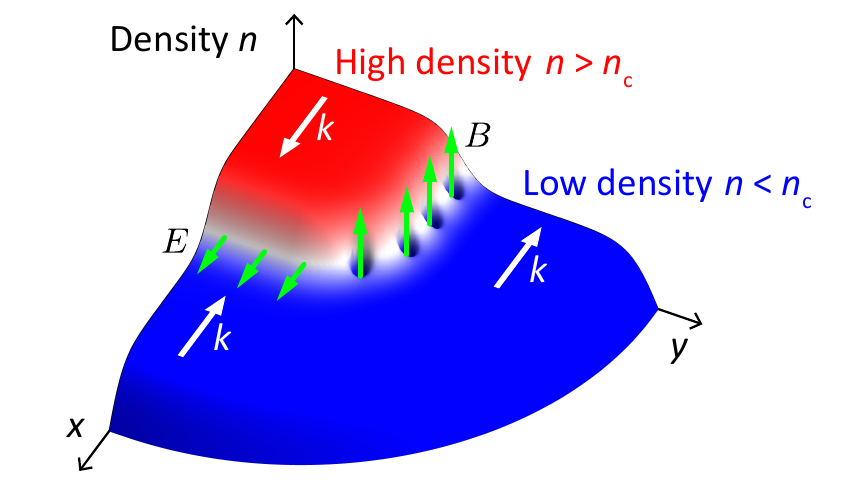}
\caption{\textbf{Bose-Einstein condensate with density dependent gauge field.}
We illustrate a condensate with inhomogenous density profile subject to a density-dependent gauge field $\mathcal{A}$, which changes sign when the density $n$ exceeds a critical value $n_c$.
The high density (red) and low density (blue) regions of the condensate form domains with distinct momenta $k =  k^*$ and $- k^*$ in the $x-$direction (white arrows), respectively.
% In the low density region (blue), the condensate momentum $p = A$ is negative along the $x-$direction.
% In the high density region (red), the condensate momentum is positive.
Along the domain wall (white) parallel to the gauge field, an array of vortices form as a consequence of phase continuity, which is a manifestation of the effective magnetic field $B \propto \partial_y n$.
% The condensate momentum thus depends on the local density, and has opposite directions in the low (blue) and high (red) density regions.
% The density profile of a BEC with density dependent gauge field across a density step is shown. When a BEC is subject to a gauge field that is slowly varying in space, its phase gradient follows the gauge field locally. 
% When the gauge field is density dependent, at the interface between two regions of different densities in the BEC, there will be magnetic field if the boundary is parallel to the gauge field. At the same time, the difference in phase gradient between the two regions results in an array of vortices along the same boundary, as a consequence of phase continuity.
% Because the difference in phase gradient accumulates to $2\pi$ over the distance $d$ between two adjacent vortices, the distance is inversely proportional to the change in gauge field $\Delta A$, $d = 2\pi/\Delta A$. 
% Interestingly, each vortex corresponds to exactly one quanta of magnetic flux. 
% Boundaries perpendicular to the gauge field, on the other hand, carries no magnetic field and no vortex.
On the other hand, dynamics of the condensate density can induce an effective electric field $E \propto \partial_t n$.
% The fields $E$ and $B$ (green arrows) can exert Lorentz force on the condensate and the domain wall.
}
\label{fig1}
\end{figure}

In this work, we quantum simulate a Bose-Einstein condensate (BEC) subject to a density-dependent gauge field, which is described by the energy functional
\begin{equation}\label{hamiltonian}
    H = \frac{1}{2m^*}|(\mathbf{p} -  \mathcal{A})\psi|^2  + \frac{1}{2} g|\psi|^4,
\end{equation}
where $\psi$ is the condensate wavefunction, $\mathbf{p}$ is the momentum operator, $m^*$ is the mass of the particle, $ \mathcal{A}$ is the density-dependent gauge field, and $g$ is the interaction strength.
We engineer a gauge field that takes one of two values according to the density $n = |\psi|^2$,
\begin{equation}\label{DDG}
\mathcal{A} = \hbar k^* ~\textrm{sign} (n-n_c) \hat x,
\end{equation}
where $k^*>0$ is a constant, $\textrm{sign}(x) = x/|x|$ is the sign function and $\hbar$ is the reduced Planck constant.
The gauge field is along the $+\hat x$ direction when the density exceeds the critical value $n_c$, and along $-\hat x$ at lower densities, see Fig.~1. 
We observe the formation of stable domain walls in the BEC, and extract an effective  charge-to-mass ratio of these topological defects from their dynamical response to the gauge field.
% We observe the formation of stable domain walls in the BEC, and study their dynamical response to the gauge field, from which we extract an effective charge-to-mass ratio of these topological defects.
% With this density-dependent gauge field, the kinetic energy of the BEC is
% \begin{equation}\label{dispersion}
% \epsilon_k = \frac{\hbar^2}{2m_x}(k_x-k^*~ \textrm{sign} (n-n_c))^2+\frac{\hbar^2k_y^2}{2m_y},
% \end{equation}
% where $k^* = A/\hbar$ and $m_{x,y}$ is the effective mass in the $x-$ and $y-$ direction. 
% (We observe domain wall and observe its response.... Summary 2 sentence)

\begin{figure*}
\includegraphics[width=129mm]{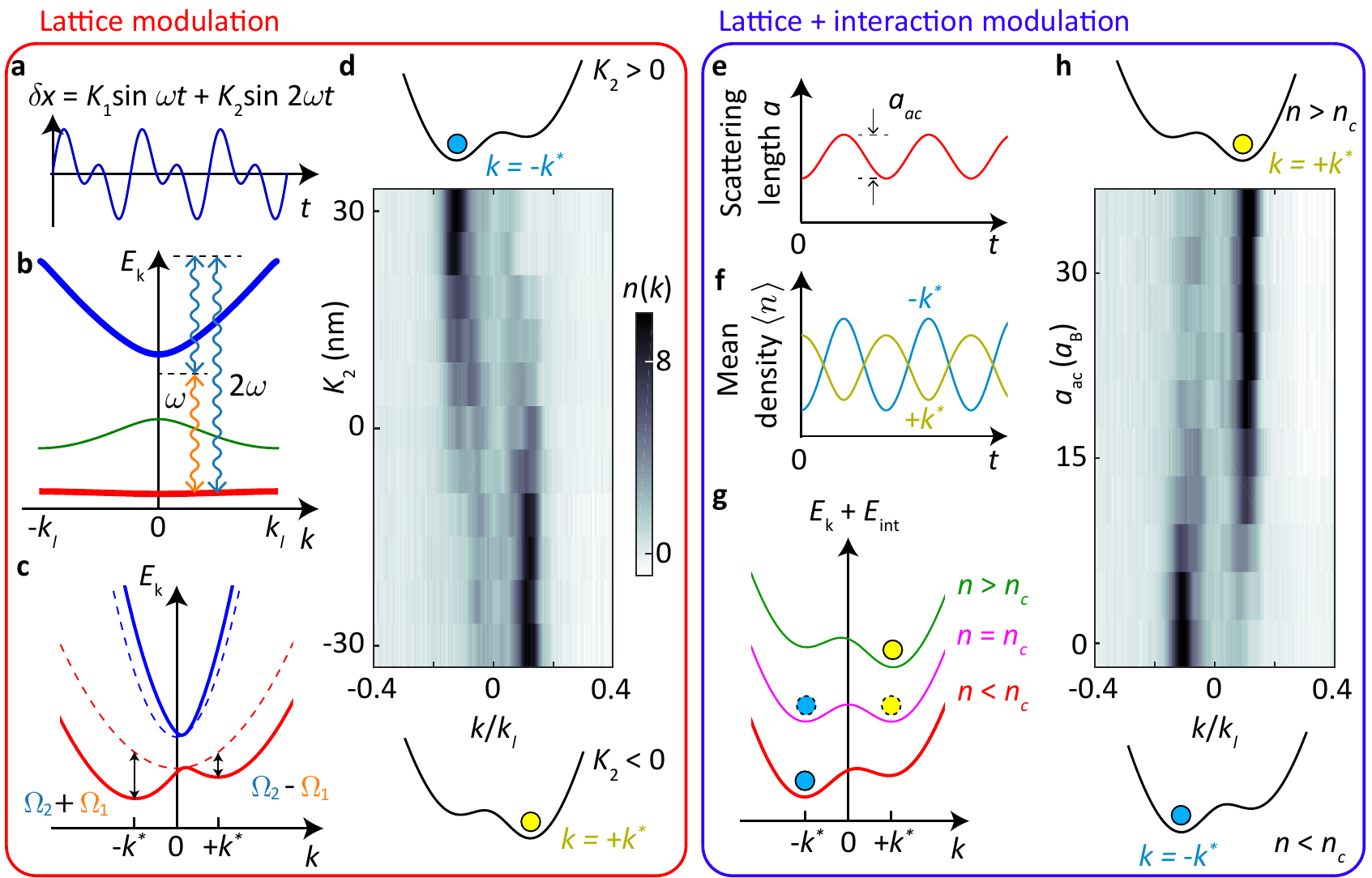}
\caption{\textbf{Creation of static (a-d) and density-dependent (e-h) gauge fields.} 
% \textbf{(a)-(d)} We create a static gauge field $A_s$ in a double well dispersion by two frequency lattice shaking.
\textbf{(a)} We periodically translate the 1D optical lattice by $\delta x = K_1 \sin \omega t + K_2 \sin 2\omega t$ with $K_1 = 21~$nm and variable $K_2$.
\textbf{(b)} The frequency $\omega$ is slightly red detuned from the transition between the ground (red) and the second excited band (blue). The first excited band (green) is only weakly coupled. Here $k_l = \pi/(532~\textrm{nm})$. The shaking introduces a direct coupling $\Omega_1$ (orange arrow) and a Raman coupling $\Omega_2$ (blue arrow). \textbf{(c)}
% \textbf{(c)} The direct coupling $\Omega_1$ has odd parity and the Raman coupling $\Omega_2$ has even parity.
In the Floquet picture, the two couplings constructively (destructively) interfere for positive (negative) $k$ when $K_2>0$.
The couplings hybridize the bare bands (dashed lines), and the resulting ground band (red line) forms a tilted double well with minima at $k \approx \pm k^* = \pm 0.15 k_l$.
% The hybridized bands (red and blue) are pushed further from the bare bands (dashed lines) for $k<0$, and the ground band forms a tilted double well.
\textbf{(d)} Time-of-flight images show a jump of the BEC momentum when $K_2$ flips sign.
See illustrations for the dispersions with $K_2>0$ and $K_2<0$.
% The observation is consistent with the BEC residing in the minimum of the dispersion subject to the static gauge field $A_s$, see illustrations.
% The measured momentum of the BEC depends on the shaking amplitude $K_2$. 
The 1D momentum distribution $n(k)$ is normalized over the first Brillouin zone.
% The $\omega$ shaking amplitude is $K_1 = 21~\textrm{nm}$.
% We use $K_2 = 23~\textrm{nm}$ (red dashed line) for the rest of this work.
% \textbf{(e) - (h)} We create a density dependent gauge field by modulating the scattering length $a$ synchronously with the micromotion from lattice shaking.
\textbf{(e)} The scattering length $a$ is modulated at frequency $\omega$.
\textbf{(f)} The micromotion of the atomic density $\langle n \rangle$ at  $k = \mp k^*$ oscillates in and out of phase with the scattering length modulation.
This results in a higher interaction energy for $k = -k^*$ than for $k = + k^*$.
\textbf{(g)}
Combining both modulations yields a dispersion whose minimum position depends on the density as $k = k^* \textrm{sign} (n-n_c)$.
% At low densities, the BEC prefers the $-k^*$ state, whereas above a critical density $n_c$, the BEC prefers the $+k^*$ state.
\textbf{(h)} The momentum distribution of the BEC displays a jump when $a_{ac}$ exceeds $9~a_\mathrm{B}$.
See illustrations for the dispersions with $n>n_c$ and $n<n_c$.
% We measure the momentum space distribution at various interaction modulation amplitudes $a_{ac}$. 
% Here we use parameters $K_1 = 21~\textrm{nm}$, $K_2 = 23~\textrm{nm}$.
}% Continued caption
\end{figure*}

% (Fig.~1) 
In the BEC described by Eq.~\eqref{hamiltonian}, the local phase gradient of the ground state wavefunction follows the gauge field, $\partial_x \phi = k^*~ \textrm{sign} (n-n_c)$, in order to minimize the kinetic energy. 
% In terms of phase gradient, or equivalently local momentum, 
The condensate can support two types of domains with momentum $k = + k^*$ for density $n$ exceeding the critical value $n_c$ and momentum $k=-k^*$ for lower density $n<n_c$. 
The density dependent magnetic field $\textbf B = \nabla \times \mathcal{A} = -2\hbar k^* \delta(n-n_c)\partial_y n \hat z$ is concentrated on domain walls parallel to the gauge field. 
% At the same time, since the two domains have different phase gradients, the phase winding requires an array of vortices along the domain wall. 
% The difference in phase gradient accumulates to $2\pi$ over the separation $d$ between adjacent vortices, $d = \pi/k^*$. Interestingly, this implies that each vortex corresponds to exactly one magnetic flux quantum.
On the other hand, dynamics of the density generates an electric field $\textbf E = -\partial_t  \mathcal{A} = -2\hbar k^* \delta(n-n_c)\partial_t n  \hat x$. 
The electromagnetic fields $E$ and $B$ can induce Lorentz force on the atoms, simulating charged particles in the gauge field.
% The Lorentz force exerted by the electromagnetic fields $E$ and $B$, in turn, can drive further dynamics in the BEC and of the domain wall.

In our experiment, we load a nearly pure BEC of around 40,000 $^{133}$Cs atoms into a one-dimensional (1D) optical lattice along the $x-$direction with an additional weak harmonic confinement in the $x-y$ plane at the radial trap frequency $2\pi \times 8$~Hz and a tight vertical confinement at trap frequency $2\pi \times 223$~Hz. 
The condensate remains in the 3D regime, with a chemical potential $2\pi \times 170$~Hz. Using Floquet engineering \cite{Eckardt2017}, we realize the gauge field in Eq.~\eqref{DDG} by generating a tilted double well dispersion $\epsilon_k$ along the lattice direction, where the energy offset of the two wells depends on the density of the sample. The dispersion can be modeled by
\begin{equation}\label{dispersion}
\epsilon_k = \alpha (k^2 - {k^*}^2)^2 - \frac{\hbar}{m^*} kA(n).
\end{equation}
Here $k$ is the wavenumber, $\alpha$ and $k^*$ can be controlled by lattice shaking along the $x-$direction, $m^*$ is the effective mass near $k = \pm k^*$, and the gauge field $A = A_s + A_d(n)$ contains the static and density-dependent contributions $A_s$ and $A_d(n)$, respectively, which we generate from a synchronous modulations of the lattice potential and the interatomic interaction, respectively \cite{Clark2018}, see Fig.~2.

%Near the bottom, the dispersion is then well described by \eqref{dispersion}, neglecting displacement of the well minima by the tilt. 

\begin{figure*}
\includegraphics[width=129mm]{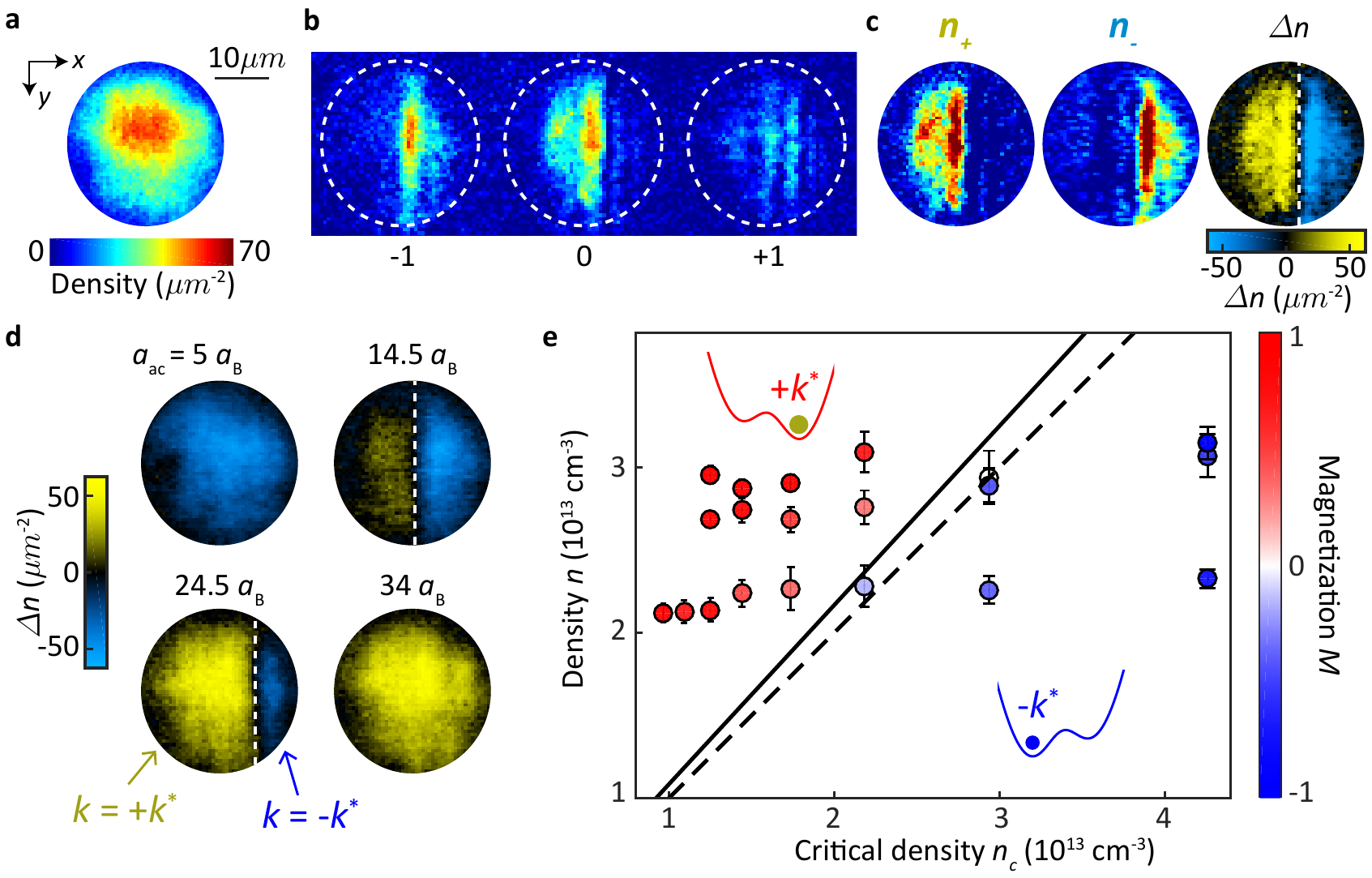}
\caption{\textbf{Domains and domain walls in the presence of density-dependent gauge field.}
\textbf{(a)} \textit{In situ} image of the BEC in a harmonic trap shows a nonuniform density profile. 
\textbf{(b)} The BEC is Bragg diffracted by the lattice after a 6~ms time-of-flight expansion. Atomic populations in $k = \pm k^*$ states are transferred to different Bragg orders.
Here a single shot image is shown.
\textbf{(c)} From the image we reconstruct the density profiles $n_{\pm}(\mathbf{r})$ of the $\pm k^*$ domains. The difference $\Delta n = n_+ - n_-$ reveals the domain structure, and $\Delta n = 0$ indicates the domain wall (white dashed line).
% We plot the same single shot example as in (b).
\textbf{(d)} Examples of the domain structure are shown at various modulation amplitudes $a_{ac}$. 
% For intermediate values of $a_{ac}$ where the condensate density is comparable to $n_c$, we observe two domains separated by a domain wall.
Each image is an average over 15 realizations.
\textbf{(e)} The magnetization $M$ near the center of the cloud is compared for different critical density $n_c$ and atomic density $n$. The dotted line indicates the predicted location for $M=0$, with $n = n_c = \epsilon/g_{ac}$ and $\epsilon = h\times 21.5$~Hz.
Experimental fit (solid line) yields  $\epsilon_\textrm{exp} = h\times 23(1)$~Hz. See supplement for details.
% To vary the density $n$, we repeat the experiment at three different total particle numbers $N = 4.8, 3.6, 2.5 \times 10^4$. 
Each data point is an average of 15 samples. Error bars denote one standard deviation.}% Continued caption
\end{figure*}

We modulate the lattice position $\delta x$ in time $t$ at two frequencies according to $\delta x(t) = K_1 \sin \omega t + K_2 \sin 2\omega t$, see Fig.~2(a), where the modulation amplitude $K_1$ determines $\alpha$ and $k^*$ of the double well dispersion, and the amplitude $K_2$ imbalances the two minima \cite{Struck2012}.
% This creates an imbalanced double well dispersion that favors the minimum $-k^*$.
The fundamental frequency $\omega$ is red detuned to the second excited band of the lattice at zero momentum, see Fig.~2(b) and supplement. 
The shaking induces a direct single photon coupling at frequency $\omega$ and coupling strength $\Omega_1$, as well as a Raman coupling involving both an $\omega$ photon and a $2\omega$ photon with coupling strength $\Omega_2$.
% The shaking results in two couplings between the ground and excited bands. One is the direct single photon coupling at frequency $\omega$ with a coupling strength $\Omega_1$. The other is the Raman coupling by absorbing a $2\omega$ photon and emitting an $\omega$ photon, with a coupling strength $\Omega_2$. 
The direct coupling $\Omega_1$ has an odd parity that only mixes states with non-zero momentum $k\neq 0$, essential for the creation of the double well dispersion, see Fig.~2(c). On the other hand, the Raman coupling $\Omega_2$ has an even parity. 
The interference of the two couplings $\Omega_1$ and $\Omega_2$ with opposite parities results in the imbalance of the two dispersion minima.
% The interference between the couplings $\Omega_1$ and $\Omega_2$ with opposite parities results in the imbalance of the two wells. 
%As a result, the two couplings constructively interfere at $k<0$, and destructively interfere at $k>0$. The hybridized ground band is thus pushed down further at $k<0$, resulting in a double well that favors $-k^*$. 
We control the imbalance in our experiment with the amplitude of the second harmonic modulation $K_2$, which results in a static gauge field $A_s \propto -K_2$. See supplement for details.

The static gauge field $A_s$ manifests in the momentum distribution of the BEC. 
% The momentum distribution is obtained by letting the cloud expand for quarter period in a weaker harmonic trap, a technique known as focused time-of-flight.
Based on the focused time-of-flight method \cite{Shvarchuck2002}, we see that the condensate momentum indeed takes on values $ k = \pm k^*$ depending on the sign of $K_2$, see Fig.~2(d). For the rest of this work, we choose $K_2 = 23~$nm, which imbalances the two wells by $h \times 3~$Hz.

The density dependent part of the gauge field $A_d$ is created by modulating the scattering length $a$ with an external magnetic field \cite{Clark2018} at the same fundamental frequency as the lattice shaking $a(t) = a_{dc} - \frac{1}{2}a_{ac} \cos \omega t$, see Fig.~2(e), where $a_{dc} = 50~a_\mathrm{B}$ and $a_{ac}$ are the mean scattering length and the amplitude of the modulation, respectively, and $a_\mathrm{B}$ is the Bohr radius. 
% We control the scattering length by tuning the magnetic field near a Feshbach resonance \cite{Chin2010}.
To understand the density dependence of the gauge field, we note that the atoms in the $k = \pm k^*$ states acquire a time dependent micromotion from the lattice shaking. 
Within a Floquet cycle, the atomic density of the two states $k = \pm k^*$ oscillates at frequency $\omega$ with opposite phase \cite{Clark2018}, see Fig.~2(f).
% This leads the atomic density to oscillate at frequency $\omega$ with opposite phase for the two states within a Floquet cycle \cite{Clark2018}, see Fig.~2(f).
% The micromotion leads to a density variation which is out of phase for atoms in the $k = \pm k^*$ states within a Floquet cycle.
We modulate the scattering length in phase with the atomic density in the state $k = -k^*$, which raises the time-averaged interaction energy for $k = -k^*$ and lowers that for $k = +k^*$.
% We modulate the interaction in phase with the density modulation of the atoms in the momentum state $k = -k^*$, and thus out of phase with the atoms with $k = +k^*$.
% \textcolor{red}{[CONTINUE HERE]}
% % The momentum state $k = + k^*$ exhibits opposite oscillations in density.
% % The micromotion of the states at the two momentum minima $\pm k^*$ exhibits opposite oscillations in density $\langle n^2 \rangle$, in phase or out of phase with the interaction modulation. 
% As a result, the time-averaged interaction energy is higher for $k=-k^*$, and lower for $k=+k^*$. 
This results in a coupling between the density and momentum, favoring the $k = +k^*$ state.
The coupling gives the density dependent part of the gauge field $A_d = \eta g_{ac} n$, where $g_{ac} = 4\pi \hbar^2 a_{ac}/m_0$ is the AC coupling constant, $m_0$ is the mass of the cesium atom and $\eta$ can be calculated from the micromotion, see supplement.

Combining the lattice and interaction modulations, we can write the resulting gauge field as
\begin{equation} \label{DDGexp}
    A = A_s + A_d(n) = \eta g_{ac}(n-n_c),
\end{equation}
where the critical density $n_c$, at which the gauge field switches sign, is given by
\begin{equation}\label{nc}
n_c = \epsilon/g_{ac},
\end{equation}
and $\epsilon = -A_s/\eta$.
When the atomic density exceeds the critical density $n_c$, the dispersion minimum switches from $k = -k^*$ to $+k^*$.
% When atoms are condensed with the density-dependent dispersion Eq.~\eqref{dispersion}, their momentum switches from $k = -k^*$ to $+k^*$ when the density exceeds the critical value $n_c$. 
% When we load the BEC into the dispersion Eq.~\eqref{dispersion}, the condensate minimizes its kinetic energy. 
% As the local density exceeds the critical density $n_c$, the local condensate momentum switches from $k = -k^*$ to $k = +k^*$.
For a BEC residing at the lowest momentum state, its momentum also changes sign when the density exceeds the critical value, see Fig.~2(g). 
Thus the BEC can be effectively described by the energy functional Eq.~\eqref{hamiltonian} with the gauge field $\mathcal{A}$ in Eq.~\eqref{DDG} that has a step function dependence on the density.
% Across the critical density $n_c$, the density dependent gauge field $A(n)$ changes sign, as illustrated in Fig.~2(g). 
% As a result, we effectively realize a gauge field with a step function dependence on density in Eq.~\eqref{DDG}.
% This constitutes the desired density dependent gauge field \eqref{DDG}, and the critical density $n_c$ is determined by where the single particle imbalance $\Delta E$ between the double well is balanced by the difference in mean field interaction energy, 
% \begin{equation}\label{nc}
%     A_d = \gamma \g_{ac} n,
% \end{equation}
% where $\gamma \propto a_{ac}$ is the strength of the interaction modulation (see supplement).

To demonstrate the effect of the density-dependent gauge field, we measure the condensate momentum in the presence of both lattice and interaction modulations. 
We find that the condensate momentum indeed changes sign from $k = -k^*$ to $+k^*$ at $a_{ac} = 9~a_\mathrm{B}$, where the critical density $n_c$ is comparable to the density of the sample, see Fig.~2(h).
Our observation is consistent with the dispersion $\epsilon_k$ in Eq.~\eqref{dispersion} with the density-dependent gauge field $A(n)$ in Eq.~\eqref{DDGexp}.
% From time-of-flight measurement, we confirm that the condensate momentum changes sign when modulation amplitude exceeds a critical value $a_{ac} = 9~a_\mathrm{B}$, at which the critical density $n_c$ is comparable to the density of the sample, see Fig.~2(h). 
% Note that here the density $n$ is constant, and the critical density $n_c$ is tuned by varying the modulation amplitude $a_{ac}$.

% [Explain how this connects to Eq.~1 and 2: BEC with this single particle dispersion has effective hamiltonian in 1 and 2] When the BEC is loaded into this dispersion, it condenses into +k or -k depending on the local density. As density exceeds... condensate momentum switches from ...
% As a result, the BEC is effectively described by 1 and 2...

In a trapped gas, where the condensate has non-uniform density, see Fig.~3(a), we expect the condensate momentum to develop spatial structures in the presence of the density-dependent gauge field.
% Density dependent gauge field can lead to spatial structures in the ground state condensate wavefunction with non-uniform density.
In the following, we investigate the formation and dynamics of domains with different momentum in the condensate.
% we investigate the dependence of local momentum pseudo-spin on the local density, from the domain distribution in a harmonically trapped cloud (Fig.~3(a)), after slowly ramping on the density dependent gauge field.

Starting with a regular BEC in a stationary 1D lattice, we slowly ramp up the lattice and interaction modulations over 300~ms. At the end of the ramp, the dispersion has two minima at $k = \pm k^*$ around which the effective mass is $m^* = 0.7 m_0$.
The BEC has a $1/e$ lifetime of 700~ms under the driving.
We measure the spatial distribution $n_\pm (\textbf{r})$ of the atoms in the $k = \pm k^*$ states by first transferring the population in the two states to different Brillouin zones, followed by a short time-of-flight which maps the population to different Bragg orders \cite{Clark2016}, see Fig.~3(b,c) and supplement.
% We reconstruct the domain distribution by exciting the $\pm k^*$ states to different Bragg orders using a short shaped pulse of lattice shaking, followed by a short 6~ms time-of-flight to separate the Bragg peaks without losing the \textit{in situ} information \cite{Clark2016} (Fig.~3(b,c)). 
Domain structures of the condensate are revealed by the density difference $\Delta n(\textbf{r}) = n_+(\textbf{r}) - n_-(\textbf{r})$.
% From the reconstructed pseudo-spin densities $n_\pm$, we reveal the domain structure by taking the difference $\Delta n = n_+ - n_-$. 

For condensates with densities comparable to $n_c$, we frequently see regions of atoms in the same momentum state separated by domain walls, see Fig.~3(d).
% The domain structure is not completely determined by the local density (Fig.~3(d)), but instead distorted by the domain walls, which are metastable topological defects \cite{Parker2013}. 
The formation of domains results from effective ferromagnetic interactions between the $+k^*$ and $-k^*$ states \cite{Parker2013}.
In most cases, a single domain wall forms perpendicular to the lattice direction.
We do not observe parallel domain walls with the predicted vortex arrays, likely due to their higher energy cost under our conditions.
% likely because with our parameters the parallel domain walls with vortex arrays cost more energy. As a result, the upper and lower wings of the cloud with low density share the same pseudo-spin as the high density center. 
In addition, we see that the left (right) side of the condensate tends to occupy rightward (leftward) momentum, see Fig.~3(d), which we attribute to the shrinkage of the cloud during the ramp that preferentially pulls atoms towards the center. See supplement for details.
% In addition, as the dispersion transitions from parabolic to double-well, the cloud slightly shrinks, due to a combination of particle loss and reduction of quantum pressure as the effective mass diverges. This shrinking biases the domain formation such that the local momentum tends to point to the center of the cloud. 
% Combined with the energy cost of the domain wall, the result is that only one domain wall forms on one side of the cloud, and the $+k^*$ domain is always on the $-x$ side.
The position of the domain wall depends on the density and the interaction modulation amplitude $a_{ac}$, providing a test of the strength of the density dependent gauge field.

% , which determines the critical density $n_c$, and at three different total particle numbers $N = 4.8, 3.6, 2.5 \times 10^4$ to change the sample density while keeping other conditions constant.
We analyze the momentum distribution in the condensate through the local magnetization defined as
\begin{equation}
M(\mathbf{r}) = \frac{n_+(\mathbf{r})-n_-(\mathbf{r})}{n_+(\mathbf{r}) + n_-(\mathbf{r})}    .
\end{equation}
A value of $M = +1$ indicates that all atoms condense in the $+k^*$ state, $M = -1$ indicates the condensate in the $-k^*$ state, and $M = 0$ indicates a domain wall. 
% A positive $M$ indicates the $+k^*$ state, whereas a negative $M$ indicates the $-k^*$ state. 

We perform the experiment with different atom numbers and modulation amplitudes $a_{ac}$.
We extract the magnetization $M$ near the center of the condensate for various atomic density $n = n_+ + n_-$ and critical density $n_c = \epsilon/g_{ac}$, see Fig.~3(e).
% We plot $M$ against the local density $n = n_+ + n_-$ and the critical density $n_c$, which is calculated from Eq.~\eqref{nc}, see Fig.~3(e). 
% Here the energy coefficient $\epsilon$ is determined from theory predictions, $\epsilon = h \times 21.5~$Hz.
% energy $\epsilon$ the coefficient $C = 3.29 \times 10^4~\textrm{m}^{-1}$ is determined from theory predictions (see supplement). 
% We repeat the experiment at various values of $a_{ac}$, which determines the critical density $n_c$, and at three different total particle numbers $N = 4.8, 3.6, 2.5 \times 10^4$ to change the sample density while keeping other conditions constant. 
We find that the local momentum indeed settles to $+k^*$ for densities exceeding $n_c$, and to $-k^*$ for $n<n_c$.
% Using the experimental data to determine the density $n_c = \epsilon/g_{ac}$ at which the transition occurs,
From the experimental data we also extract the coefficient $\epsilon$, and the result $\epsilon_\mathrm{exp} = h \times 23(1)~$Hz is in good agreement with the prediction $\epsilon = h \times 21.5~$Hz.

\begin{figure}
\includegraphics[width=86mm]{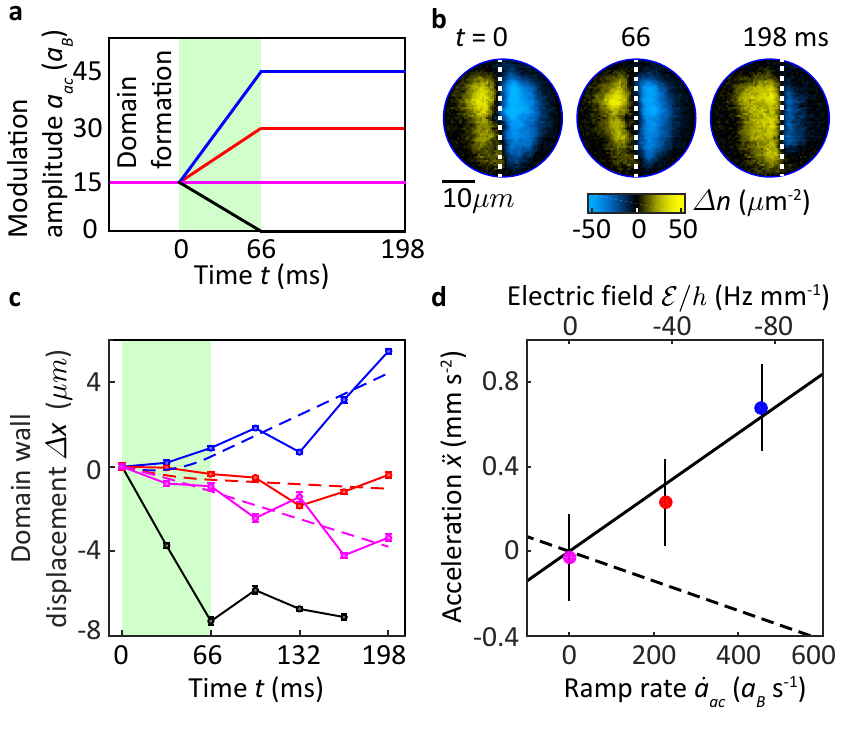}
\caption{\textbf{Dynamics of the domain wall in response to a synthetic electric field $\mathcal{E}$.}
\textbf{(a)} After forming the domains at modulation strength $a_{ac} = 15~a_\mathrm{B}$, we ramp to $a_{ac}=0$ (black), $15$ (magenta), $30$ (red) or $45~a_\mathrm{B}$ (blue) over 66~ms and hold for 132~ms. The ramp induces an electric field $\mathcal{E}\propto \dot{a}_{ac}$ (green shaded area).
Example images for the ramp to $a_{ac} = 45~a_\mathrm{B}$ are shown in \textbf{(b)}. The white dashed lines mark the positions of the domain walls.
% The Roman numerals indicate where the examples in (b) are from.
% \textbf{(b)} We show the domain structure at various times, with final modulation amplitude $a_{ac}=45~a_\mathrm{B}$.
% The domain wall (white dashed line) visibly moves to the right. 
Each image is the average of 15 samples.
Panel \textbf{(c)} shows the domain wall dynamics;
% , with final modulation amplitudes $0$ (black), 15~$a_\mathrm{B}$ (magenta), 30~$a_\mathrm{B}$ (red) and 45~$a_\mathrm{B}$ (blue). 
% The direction of motion of the domain wall clearly depends on the final modulation amplitudes.
dashed lines are fits based on Eq.~\eqref{acceleration}. The black data points are excluded from the fit because the domain wall moves out of the cloud.
% The domain wall position is determined from the average of 5 to 15 samples. 
\textbf{(d)}
The acceleration $\ddot{x}$ extracted from the fit shows a linear dependence on the ramp rate $\dot{a}_{ac}$ and the electric field $\mathcal{E}$.
% , corresponding to different effective electric field $\mathcal{E}$.
The linear fit $\ddot{x} = \beta \dot{a}_{ac}$ (black line) gives $\beta = -26(6)~\textrm{ms}^{-1}$. The prediction for bare atoms gives $\beta_\mathrm{atom} = 13~\textrm{ms}^{-1}$ (dashed line).
% The effective charge-to-mass ratio $Q/M$ of the domain wall is -2.0(5) times that of atoms. 
Error bars indicate one standard deviation.
}
\label{Fig4}
\end{figure}

The deterministic formation of domains offers an opportunity to study the domain walls as elementary objects, which is of fundamental interest to condensed matter physics \cite{Mermin1979}, high energy physics \cite{Gani2014} and cosmology \cite{Vilenkin1985}. 
% Previous theoretical analysis suggests that the domain walls at rest cost a finite amount of energy $E_0 = \frac{8}{3} \epsilon n \Lambda /k^*$, where $\Lambda$ is the area of the domain wall and $\epsilon$ is the barrier height of the double well dispersion \cite{Liu2016}. This energy is approximately $ k_B \times 1$~nK per atom under our parameters, carried by atoms near the domain wall. 
% However, the motion of the domain walls as topological defects has not been studied.
% \textcolor{red}{[CONTINUE HERE]}
We introduce a phenomenological model that describes the domain wall as an elementary excitation with charge $Q$ and mass $M$ interacting with the gauge field $\mathbf{A}$, with energy
\begin{equation}
    E = \sigma \Lambda+\frac{(\mathbf{P} - Q \mathbf{A})^2}{2M},
\end{equation}
where $\mathbf{P} = M\mathbf{v}+Q\mathbf{A}$ and $\mathbf{v}$ are the canonical momentum and the velocity of the domain wall, $\Lambda$ is the area of the domain wall, the surface tension $\sigma =  \frac{8}{3} \varepsilon n /k^*$ is calculated in \cite{Liu2016}, and $\varepsilon $ is the barrier height of the double well dispersion. 
For our parameters $\varepsilon = h \times 4~$Hz and the rest energy of the domain wall $\sigma \Lambda$ is $\approx k_B \times 1$~nK per atom in the domain wall. 

For our observed domain walls perpendicular to the lattice along the $x-$axis, their motion is restricted to the same direction.
The dynamics is driven by the Lorentz force with only the electric field in the $x-$direction $\mathcal{E} = - \partial_t A$, with $A$ given in Eq.~\eqref{DDGexp}.
% A time dependent gauge field creates an electric field $\mathcal{E} = - \partial_t A$ that drives the domain wall according to the Lorentz force law $M \ddot{x} = Q \mathcal{E}$. 
We derive
\begin{equation}\label{acceleration}
\begin{split}
\ddot{x} &= \frac{Q}{M} \mathcal{E}\\
\mathcal{E} &= -  \partial_t\eta (g_{ac}  n-\epsilon).
\end{split}
\end{equation}
% where $Q/M$ is the charge to mass ratio of the domain wall.
    % \ddot{x} = -\frac{Q}{M} \partial_t A = -\frac{Q}{M} \eta \partial_t (g_{ac}  n),
    
To study the dynamical response of the domain wall to the electric field $\mathcal{E}$, we ramp the density dependent gauge field and monitor the motion of the domain wall.
After preparing one domain wall in the BEC at the modulation strength $a_{ac}=15~a_\mathrm{B}$, we ramp $a_{ac}$ to different values over 66~ms, which induces an electric field $\mathcal{E}$. We then hold for another 132~ms during which the domain wall can freely propagate, see Fig.~4(a).
% over 66~ms and hold for another 132~ms, see Fig.~4(a).
% We find that no more domain walls are created in the process, and the dynamics of the system is well described by the motion of a single domain wall in the lattice direction, see Fig.~4(b). 

We observe that the domain wall moves in the lattice direction in response to the ramp, see Fig.~4(b,c), consistent with the direction of the electric field. 
% We extract the location of the domain wall by fitting the zero crossing of the domain density $\Delta n$, see supplement.
% From the domain configurations we extract the positions of the domain walls by fitting the zero crossing of the domain density $\Delta n$, integrated over the $y-$direction. 
% The domain wall motion depends on the ramp rate $\dot{a}_{ac}$, and 
The motion persists in the same direction after the ramp stops. 
From Eq.~\eqref{acceleration} we expect that the domain wall accelerates during the ramp $\ddot{x} = \beta \dot{a}_{ac}$, where $\beta \propto Q/M$, and maintains a constant velocity during the hold time. (The atomic density $n$ remains almost a constant to within 20\% during the dynamics, and $\eta$ and $\epsilon$ are constants.)
We fit the domain wall trajectories to extract the acceleration $\ddot{x}$, which indeed shows a linear dependence on the ramp rate $\dot{a}_{ac}$, see Fig.~4(d). 
% The fit gives a proportionality constant $\beta = 1.4(3)~\mu m ~{a_\mathrm{B}}^{-1}\textrm{s}^{-1}$, whereas the predicted proportionality constant for atoms is $\beta_\mathrm{atom} = 0.7~\mu m ~{a_\mathrm{B}}^{-1}\textrm{s}^{-1}$, see supplement. 
% We thus obtain that the charge-to-mass ratio of the domain wall $\xi = 2.8(7) ~{m_\mathrm{Cs}}^{-1}$ is 2.0(5) times that of the atoms $\xi_{atom} = 1.4~ {m_\mathrm{Cs}}^{-1}$.
From the linear fit we extract the charge-to-mass ratio of the domain wall to be $Q/M = -2.8(7) ~{m_0}^{-1}$, where $m_0$ is the mass of a cesium atom.

% (Explain how to interpret, why compare with atoms, why is the deviation interesting)
Our measurements present an interesting result where the topological defect in the BEC with density-dependent gauge field behaves very differently from the bare atoms.
For bare atoms in the condensate with the same microscopic dispersion as in Eq.~\eqref{hamiltonian}, the charge-to-mass ratio is $1/m^* = 1.4~ {m_0}^{-1}$.
% (suggest defect moves in opposite direction)
% which should be compared with that of the atoms $1/m^* = 1.4~ {m_0}^{-1}$.
This suggests that the electric field propels the domain wall in the opposite direction compared to the bare atoms at 2.0(5) times the acceleration.
% This suggests that the domain wall has 2.0(5) times higher sensitivity to the electric field than the atoms, and responses in the opposite direction.
% This suggests that the charge-to-mass ratio of the domain wall is -2.0(5) times larger than the atoms.
% This difference highlights the different response of the topological defect to external drive compared to the constituent particles, which is a common phenomenon in condensed matter physics.
Notably, the direction of domain wall motion is consistent with the condensate relaxing to the momentum state with lower energy.
A quantitative understanding of the different responses between the domain wall and the bare atoms demands further theoretical and experimental investigation.

% Understanding the dynamical properties of topological defects is of fundamental interest, as the defects play an important role in many areas of condensed matter \cite{Hughes2017} and high energy physics \cite{Brandenberger1993}.
% , such as magnetic materials, nematic liquid crystals and cosmic strings. 
% Our platform offers the prospect to model the topological defects as quasi-particles and study their interactions in a quantum many-body system.

In summary, we demonstrate deterministic creation of domain walls in a BEC with density-dependent gauge field, created by simultaneous modulations of the lattice potential and the interaction strength. The domain walls remain stable in the BEC and behave like elementary excitations. Their dynamical response to the gauge field is observed to be drastically different from the bare atoms.
Our work offers promising prospects to simulate the dynamics and interactions of topological defects such as domain walls and vortex lines in quantum many-body systems with dynamical gauge fields.

We thank E. Mueller for helpful discussions, and K. Patel for carefully reading the manuscript. This work is
supported by the National Science Foundation (NSF) grant
no. PHY-1806733, NSF QLCI-HQAN no. 2016136, the Army Research Office STIR grant W911NF2110108, and the U.S. Department of Energy, Office of Basic Energy
Sciences, under contract number de-sc0019216.

\bibliographystyle{apsrev4-1}
\bibliography{scibib}

% \begin{thebibliography}{10}

% \bibitem{betzig2006imaging}
% E.~Betzig, G.~H. Patterson, R.~Sougrat, O.~W. Lindwasser, S.~Olenych, J.~S. Bonefacino, M.~W. Davidson, J.~Lippincott-Schwartz, H.~F. Hess, {\it Science} {\bf 313}, 1642--1645 (2006).

% \bibitem{betzig2006imaging}
% E.~Betzig, G.~H. Patterson, R.~Sougrat, O.~W. Lindwasser, S.~Olenych, J.~S. Bonefacino, M.~W. Davidson, J.~Lippincott-Schwartz, H.~F. Hess, {\it Science} {\bf 313}, 1642--1645 (2006).
% \end{thebibliography}

\clearpage
\widetext
\setcounter{equation}{0}
\setcounter{figure}{0}
\setcounter{table}{0}
\setcounter{page}{1}
\makeatletter
\renewcommand{\theequation}{S\arabic{equation}}
\renewcommand{\thefigure}{S\arabic{figure}}
\renewcommand{\thetable}{S\arabic{table}}

\noindent \textbf{Supplementary Material}\\

\section{Floquet engineering of the gauge fields $A_s$ and $A_d$}

An atom in our shaken optical lattice evolves according to the following Hamiltonian,
\[
H = \frac{p^2}{2m} + \frac{U}{2} \cos k_0 (x - \delta x),
\]
where $p$ is the 3D momentum of the atom, $U$ is the lattice depth, $ k_0$ is the lattice wavenumber, $\delta x = K_1 \sin \omega t + K_2 \sin 2\omega t$ is the lattice displacement.
On the single particle level, the dynamics in the $y-$ and $z-$ direction are decoupled, and we focus on the $x-$ direction. 
The time dependent Hamiltonian has discrete translational symmetry of the lattice, and the Hamiltonian separates for different quasi-momentum quantum numbers $k$ as $H = \bigotimes_k  H(k)$.
We numerically calculate the dispersion of the Floquet bands by diagonalizing the Floquet operator $U_F(k) = e^{-i \int_0^T H(k) dt}$ in momentum space, including the first 15 bands in the Hilbert space, and Trotterizing the time evolution into 100 steps. 

The operator is diagonalized as $U_F(k) = \sum_j e^{-i \epsilon_j(k) T/\hbar}|\psi_j(k)\rangle \langle\psi_j(k)|$. The eigenvalues $\epsilon_j(k)$ are the quasi-energies, giving the effective dispersion of the hybridized bands. The eigenvectors contain the micromotion of the Floquet eigenstates $|\Psi_j(k,t)\rangle = e^{-i \int_0^t H(k)d\tau}|\psi_j(k)\rangle$, from which we calculate the micromotion of the density $\langle n (t) \rangle = \int |\Psi_j (x,t)|^4 dx$ shown in Fig.~2(f).

The scattering length is modulated as $a(t) = a_{dc} - \frac{1}{2} a_{ac} \cos \omega t$. The time averaged interaction energy (chemical potential) is $E_\mathrm{int} = \frac{N}{V} \frac{1}{T}\frac{4\pi \hbar^2}{m_0}\int \langle n(t) \rangle  a(t) dt$, for $N$ atoms in volume $V$, corresponding to experimentally measured atomic density $N/V$, which is averaged over length scales larger than the lattice constant.

Comparing the interaction energy $E_\mathrm{int}$ for $k = \pm k^*$ states, we obtain the factor $\eta$ in the expression of the density dependent gauge field $A_d$ Eq.~\eqref{DDGexp}. This approach treats the interaction effects to zeroth order in perturbation since we neglect the deviation in density profile from the single particle eigenstates due to interactions.

Analytically we can obtain a qualitative understanding of the creation of the tilted double well dispersion from perturbation theory. Performing the Jacobi-Anger expansion on the lattice potential, we arrive at
\[
H = -\frac{\hbar^2}{2m} \partial_x^2 + \frac{U}{2} \cos  k_0 x + H_1 = H_0 + H_1,
\]
where $H_0$ describes the static lattice, and $H_1$ describes the driving,
\[
H_1 = \frac{U}{4} (e^{i k_0x} f + e^{-i k_0x} f^*),
\]
\[
f = -\frac{1}{4}(\alpha^2+\beta^2) + 2i\alpha \sin \omega t - 2 \alpha \beta \cos\omega t.
\]
Here $\alpha =  k_0 K_1$, $\beta =  k_0 K_2$, and we keep terms up to second order in $\alpha$ and $\beta$, and up to $\omega$ in frequency.

The eigenstates of $H_0$ are the Bloch waves. Consider the states $|0,k\rangle$ and $|2,k\rangle$ in the ground and second excited bands at quasimomentum $k$. Under rotating wave approximation, the effective Hamiltonian is
\[
H_{\textrm{eff}} = 
\begin{pmatrix}
E_0 & \Omega \\
\Omega^* & E_0+\Delta 
\end{pmatrix},
\]
where $E_0 = \langle 0,k|H_0|0,k \rangle$ is the bare energy of the ground band, $\Delta$ is the detuning, and the coupling is 
\[
\Omega = \alpha \Omega_- - \alpha \beta \Omega_+ .
\]
Here $\Omega_\pm = \langle 0,k|e^{i  k_0 x} \pm e^{-i k_0x} |2,k\rangle$. From here we can see that the coupling has two contributions, one is the direct coupling $\Omega_1 = \alpha \Omega_-$, the other is the Raman coupling $\Omega_2 = -\alpha \beta \Omega_+$. The parity of $\Omega_-$ is odd, and that of $\Omega_+$ is even, because the ground and second excited bands both have even parity wavefunctions.

Near $k = 0$, to first order the matrix elements depend on quasimomentum $k$ as $\Omega = \alpha \omega_0 k- \alpha \beta \omega_1 $, $E_0 = \epsilon_0 k^2$ and $\Delta = \epsilon_1 k^2 + \Delta_0$. Then the hybridized ground band dispersion is  
\[
E_g = \epsilon_0 k^2 + \frac{1}{2}\left(\epsilon_1 k^2 + \Delta_0  - \sqrt{4(\alpha \omega_0 k- \alpha \beta \omega_1)^2+(\epsilon_1 k^2 + \Delta_0)^2}\right).
\]
The dispersion has the shape of a double well because the coupling has a zero crossing near $k=0$. Since the fundamental shaking frequency is red detuned, the coupling pushes down the ground band energy. The tilt is a result of the constructive and destructive interference of $\Omega_1$ and $\Omega_2$ at positive and negative quasi-momentum, which pushes down the ground band energy more on one side than the other.
To lowest order, this tilt is given by a linear term in the dispersion $2  \alpha^2 \beta \omega_0 \omega_1k/\sqrt{4 (\alpha \beta \omega_1)^2 + \Delta_0^2}$, which effectively generates a static gauge field $A_s \propto \beta = k_0 K_2$. The sign of the gauge field depends on the phase between the $K_1$ and $K_2$ lattice modulation components.

% The dispersion has the shape of a double well because the coupling has a zero crossing near zero quasimomentum. Since the fundamental shaking frequency is red detuned, the coupling pushes down the ground band energy on the sides of the zero crossing. The tilt is a result of the constructive and descructive interference of $\Omega_1$ and $\Omega_2$ at positive and negative quasi-momentum, which pushes down the ground band energy more on one side than the other.

The numerical Floquet calculation indicates that the modulation weakly couples the ground band to the first excited band in addition to the second excited band. The coupling to the first excited band mostly contributes to a constant energy shift, and does not qualitatively change the shape of the dispersion.

\section{System preparation}
In our experiment, the optical lattice is formed by a pair of counter-propagating 1064~nm lasers, with lattice constant 532~nm. We use parameters lattice depth $U=8.9 E_R$, where $E_R = h \times 1.3~$kHz is the recoil energy, and $\omega = h \times 9091~$Hz. Under our conditions, the factor $\eta$ in Eq.~\eqref{DDGexp} is $\eta  = 0.07m^*/\hbar k^*$, where $m^* = 0.7 m_0$ and $k^* = 0.15 k_l$.
 
After loading the atoms into the 1D optical lattice with harmonic confinement formed by 1064~nm lasers, we prepare the BEC under density-dependent gauge field by slowly ramping up the modulation amplitudes. We ramp up the amplitude $K_1$ to 7~nm over 11~ms (100 oscillation periods). Since the critical shaking amplitude for the formation of double well dispersion is 14~nm (obtained from the Floquet calculation of dispersion), the effective dispersion changes very little during this time, and we ramp quickly to reduce particle loss. We then ramp up the amplitude $K_1$ to 21~nm over another 289.3~ms (2630 oscillation periods), which gives a ramp rate slow enough to suppress fluctuations from the Kibble-Zurek mechanism \cite{Clark2016} and allow for deterministic evolution of the system. The amplitudes $K_2$ and $a_{ac}$ are ramped to the final value over the first 11~ms. This ramp procedure turns on the gauge field slowly over time, and results in a roughly constant critical density $n_c$ throughout the ramp.

Although the dynamics during the ramp on of the gauge is deterministic, it is not quite adiabatic since the two momentum minima are only offset by $h\times 3~$Hz, comparable to the ramp time 300~ms, and we do not arrive at the ground state. During the ramp fields, the cloud systematically shrinks, in part due to particle loss which reduces the chemical potential, and in part due to the reduction of quantum pressure as the dispersion crosses the critical point from parabolic to double well, during which the effective mass diverges and the quantum pressure drops to zero. Since we are in the Thomas-Fermi regime, the quantum pressure is usually negligible, but in this case its reduction is significant enough to bias the domain formation because a slow ramp across the critical point is very susceptible to any bias. We have confirmed this effect in experiments with no gauge field (balanced double well dispersion), and in numerical simulations without particle loss.

\section{Extracting the domain densities from Bragg peaks}

We extract the spatial distribution of the atoms in the $k = \pm k^*$ states following the technique in \cite{Clark2016}. At the time of detection, we switch off $K_2$ and $a_{ac}$ and ramp the modulation amplitude $K_1$ to 140~nm over 0.8~ms. This pulse of lattice shaking excites the atoms from the ground band to superposition states of excited bands at the same quasi-momentum, which have oscillating projections to each Brillouin zone. Atoms in different quasi-momentum states have different oscillations. We image the atoms at the time when the projections of $k = \pm k^*$ states are maximally different. We perform a 6~ms time-of-flight to map the Brillouin zones to Bragg diffraction orders. 

From the densities in the Bragg diffraction orders 
\[
\vec{n}(x,y) = \left(n_{-1}(x,y),n_0(x,y),n_1(x,y)\right),
\]
we fit using the ansatz that 
\[
\vec{n} = n_+ \hat{e}_+ + n_- \hat{e}_-,
\]
where the basis vectors $\hat{e}_\pm$ are calibrated by biasing the entire condensate into $k = \pm k^*$. The basis vectors $\hat{e}_\pm$ are $L^1$ normalized, as they represent density distributions of the $k = \pm k^*$ states. In the fit we impose the positivity constraint $n_\pm>0$.
% We then fit the measurement as a linear combination of contributions from $k = \pm k^*$ states, under the constraint that $n_\pm \geq 0$.

The Bragg peaks of atoms in the $k = \pm k^*$ states are shifted relative to each other during the TOF, because of the difference in quasi-momentum. We take this shift into account when reconstructing the domain densities. Additionally, this shift may cause originally disjoint domains to overlap during the TOF. The coherent domains interfere in the overlapping region, forming density waves at wavenumber $2k^*$. This effect does not significantly alter the extracted domain structure or domain wall position, and we neglect it in our analysis.

\section{Analysis of the domain structures}

Since we observe that the domain walls are mostly perpendicular to the lattice direction, in our analysis we treat the domain structures as 1D. For the analysis in Fig.~3(e) of the main text, we integrate the mean and difference of the domain densities, $n = n_+ + n_-$ and $\Delta n = n_+ - n_-$, over the $y-$direction, then select the central 10\% of the cloud. Effectively we select a central vertical stripe of the cloud. We have checked that our results are not sensitive to the chosen stripe width. From each experimental realization we calculate the magnetization $M = \Delta n / n$, and we plot the average of $n$ and $M$ for each set of modulation amplitude $a_{ac}$ and particle number $N$. We convert the 1D density to 3D density by dividing with the length scales in the $y$ and $z$ directions, $l_y$ and $l_z$. Since the chemical potential is not larger than the trap frequency in the $z-$direction, we use the length scale of the harmonic oscillator ground state $l_z = \sqrt{h/m\omega}$. We obtain the length scale $l_y = (\int n dy)^2 / \int n^2 dy$ from the measured density profiles $n$.

From the experiment data in Fig.~3(e), we extract a value of $\epsilon_\mathrm{exp}$ in Eq.~\eqref{nc} by fitting to the expression
\[
M = \tanh \frac{\ln n - \ln(\epsilon_\mathrm{exp}/ g_{ac})}{C},
\]
with each data point in Fig.~3(e) corresponding to a magnetization $M$, a density $n$, and a modulation strength $g_{ac}$.
This expression represents the relation $M = \textrm{sign}(n-\epsilon_\mathrm{exp}/g_{ac})$,
but smooths the step function by a width parameter $C$. We present our fit to experiment data in Fig.~S1.

\begin{figure}
    \includegraphics[width = 86mm]{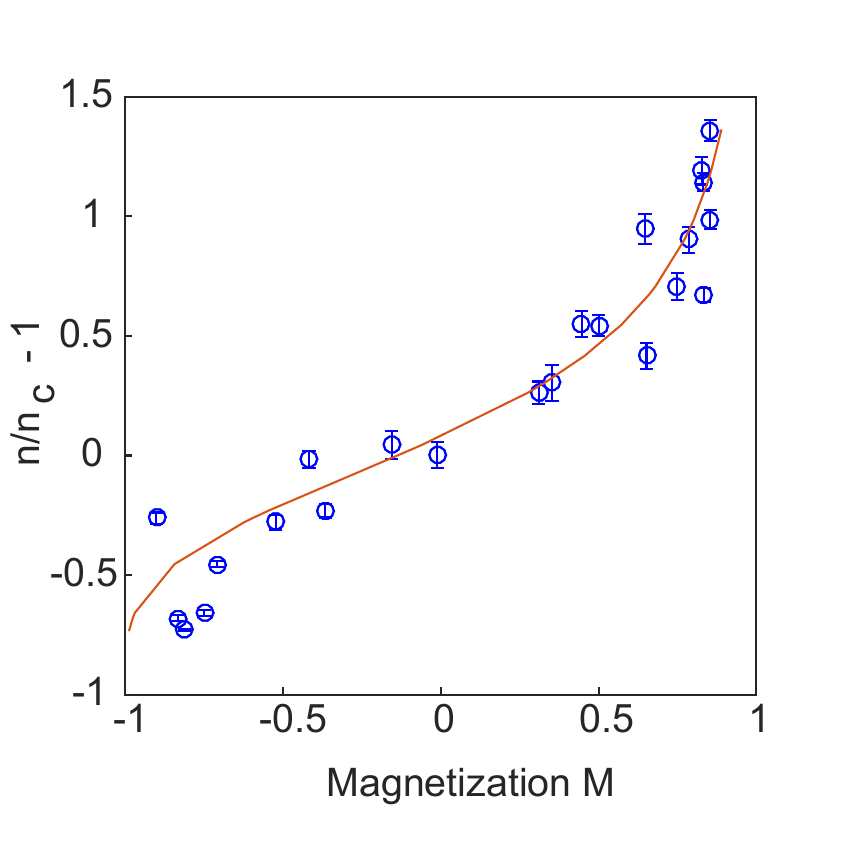}
    \caption{
    Fit to experiment data in Fig.~3(e) for the extraction of $\epsilon_\mathrm{exp}$.
    }
\end{figure}

For the analysis in Fig.~4(c), we integrate the difference of the domain densities $\Delta n$ over the $y-$direction. We then extract the position of the zero crossing of the integrated 1D domain density, by fitting a straight line to the six data points (each corresponding to a pixel in the image) around the numerical zero crossing, in order to improve accuracy. The error bars shown in Fig.~4(c) are 68\% confidence intervals of this fit.

We fit the domain wall trajectories in Fig.~4(c) by assuming a common initial velocity for all ramp rates, a constant acceleration during the ramp which is independently varied for each ramp rate, and a constant velocity after the ramp stops. The fitted initial velocity is $-17(10) ~\mu m/\mathrm{s}$, which we attribute to residual dynamics during the domain formation process.

The conversion of the ramp rate $\dot{a}_{ac}$ to the electric field $\mathcal{E}$ is derived from Eq.~\eqref{DDGexp}. We have
\[
\mathcal{E}= \frac{4\pi \hbar^2}{m_0}n\eta  \dot{a}_{ac},
\]
with density $n = 2.8\times 10^{13}~\textrm{cm}^{-3}$ from the experiment. The prediction of parameter $\beta$ for bare atoms is obtained from this relation and the charge-to-mass ratio $1/m^*$.

\end{document}